# Super-resolution for a point source better than λ/500 using positive refraction

**Juan C. Miñano[1], Ricardo Marqués[2], Juan C. González[1], Pablo Benítez[1], Vicente Delgado[2], Dejan Grabovičkić[1] and Manuel Freire[2]**

[1] Universidad Politécnica de Madrid, Cedint, Campus de Montegancedo, 28223 Madrid, Spain

[2] Departamento de Electrónica y Electromagnetismo, Facultad de Física, Universidad de Sevilla, Avda. Reina Mercedes s/n, E41012 Sevilla, Spain.

**Abstract.** Leonhardt demonstrated (2009) that the 2D Maxwell Fish Eye lens (MFE) can perfectly focus 2D Helmholtz waves of arbitrary frequency, i.e., it can perfectly transport an outward (monopole) 2D Helmholtz wave field, generated by a point source, towards a "perfect point drain" located at the corresponding image point. Moreover, a prototype with $\lambda/5$ super-resolution property for one microwave frequency has been manufactured and tested (Ma et al, 2010). However, software simulations or experimental measurements for a broad band of frequencies have not yet been reported. Here we present simulations with a non-perfect drain for a device equivalent to the MFE, called the Spherical Geodesic Waveguide (SGW), that predicts up to $\lambda/500$ super-resolution close to discrete frequencies. These frequencies are directly connected with the well-known Schumann resonance frequencies of spherical symmetric systems. Out of these frequencies, the SGW does not show super-resolution in the analysis performed.

## 1. Introduction

"Perfect imaging" stands for the capacity of an optical system to produce images with details unlimited by the wavelength of light. Perfect imaging has been theoretically demonstrated in the last decade using left-handle materials (that is, materials with negative dielectric and magnetic constants)[1][2] and super-resolution (i.e., objects resolved below the diffraction limits) based on these materials has been shown experimentally [3][4]. Unfortunately, high absorption and small (wavelength scale) source-to-image distance are inevitable in negative refraction [5].

Recently, a new possibility for perfect imaging has been proposed using a material with a positive, isotropic and gradient refractive index: the Maxwell Fish Eye (MFE) lens. It is well known that, in the Geometrical Optics framework, the MFE perfectly focuses rays emitted by an arbitrary point of space onto another (its image point). Leonhardt [6] demonstrated that the MFE lens in two dimensions (2D) perfectly focuses radiation of any frequency between the source and its image for 2D Helmholtz fields (which describes TE-polarized modes in a cylindrical MFE, i.e., in which electric field vector points orthogonal to the cross section of the cylinder); this result has also been confirmed via a different approach [7].

This "perfect focusing" stands for the capacity of an optical system to perfectly transport an outward (monopole) 2D Helmholtz wave field, generated by a point source, towards an "infinitely-well localized drain" (which we will call "perfect point drain") located at the corresponding image point. That perfect point drain must be able to absorb totally all incident radiation, with no reflection or scattering, and the field around the drain asymptotically coincides with an inward (monopole) wave.

The physical significance of a passive perfect point drain has been considered as very controversial [8]-[17]. In references [6] and [7] the perfect point drain was not physically described, but only considered as a mathematical entity. However, recently, a rigorous example of passive perfect point drain for the MFE has been found, clarifying that controversy [18]. It consists of a dissipative region whose diameter tends towards zero and whose complex permittivity $\varepsilon$ takes a specific value depending on the size of the drain as well as on the radiation frequency.

However, there is another aspect in [6] and [7] that could still be considered controversial: the perfect focusing of the MFE at a perfect drain for any frequency was proven, and perfect imaging was claimed to be inferred from it. This inference could be criticized, since the ability for perfect focusing only indicates that all the power produced by a single point source is absorbed at the perfect point drain when the latter is located at the image point, but in principle it does not provide information on how much power the drain will absorb when it is displaced to a different location. Would this perfect imaging capability give super-solution to the MFE?.

Recently two sets of experiments have been carried out to support the super-resolution capability in the MFE. In the first one, super-resolution with positive refraction has been demonstrated for the very first time at a microwave-frequency ($\lambda$=3 cm) [19][20]. In this experimental set-up, a two-dimensional MFE medium was assembled as a planar waveguide 5 mm thick with concentric layers of copper circuit board and dielectric fillers forming the desired refractive index profile of the MFE. This profile is maded only up to a diameter of 10 cm index and limited by a metallic mirror, shaping a device called MFE mirror [6][21] that has ideal properties similar to these of the MFE. Sources and drains were built as coaxial probes, inserted through the bottom plate. However, the experimental results showed that two sources with a distance of $\lambda$/5 from each other (where $\lambda$ denotes the local wavelength $\lambda = \lambda_0/n$) could be resolved with an array composed of 10 drains spaced $\lambda$/20, which surpassed the $\sim\lambda$/2.5 classical diffraction limit [19]. Results with closer sources were not reported, but it should be noted that this experiment was limited to the resolution of the array of drains.

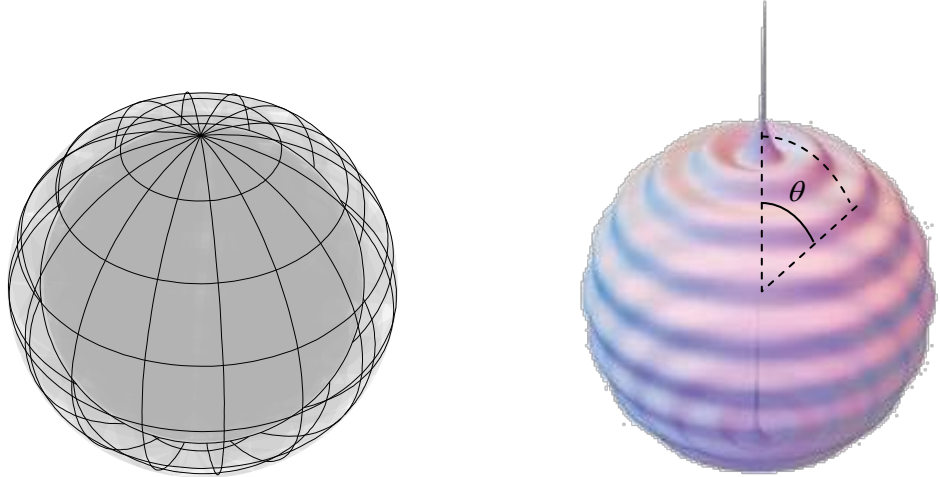

Fig. 1. Spherical Guide Wave (SGW) analyzed in this paper. The SGW is bounded by two spherical shells made of perfect conductors. On the left, dark gray is the inner metallic sphere and clear gray the outer one. On the right the electric field inside the guide for point source and perfect point drain on the opposite pole.

The second set of experiments has been carried out for the near infrared frequency ($\lambda$ = 1.55 µm), but resolution below the diffraction limit was not found [22]. In this case the planar waveguide is filled with a medium made of nano-cylinders of silicon with a different radius and separation between them, so they emulate the MFE distribution. The authors assume that the failure in the experimental demonstration is due to manufacturing flaws in the prototype. However, according to the results presented here, it seems that the reason is that super-resolution only appears close to a specific set of frequencies, known as the Schumann frequencies (see section 4). The frequency of the first experiment ($\lambda$ = 3 cm) is close to one of these particular frequencies but the second experiment ($\lambda = 1.55$ µm) is far from them.

In this paper we present the results of the super-resolution analysis for microwaves of the Spherical Geodesic Waveguide (SGW), a device suggested in [24] for perfect imaging with waves (see Fig. 1). It is obtained via transformation optics from a MFE planar waveguide. The SGW is a spherical waveguide filled with a non-magnetic material and isotropic refractive index distribution proportional to $1/r$ ($\varepsilon = (r_0/r)^2$ and $\mu = 1$), $r$ being the distance to the center of the spheres. Transformation Optics theory [25] proves that the TE-polarized electric modes of the cylindrical MFE [6] are transformed into radial-polarized modes in the SGW, so both have the same perfect focusing properties.

When the waveguide thickness is small, the variation of the refractive index within the two spherical shells can be ignored resulting in a constant refractive index within the waveguide. This is obviously very attractive from the practical point of view. The waveguide can be manufactured with just two concentric metallic spheres separated at a distance much less than the radius and constant refraction index between them. Both devices (both with and without gradient index) have been analyzed in this paper, with source and drain implemented with coaxial probes.

In section 2 the equivalence between the MFE and the spherical waveguide is reviewed. The microwave circuit model of that spherical waveguide is described in section 3. Results showing up to $\lambda/500$ super-resolution, are given in section 4. In section 5 we present a comparative analysis between the results described in section 4 and the experimental results of the prototypes presented in [19], [20] and [22]. The conclusions are given in section 6.

## 2. MFE lens and Spherical Geodesic Waveguide (SGW).

A cylindrical MFE is a lens with the following refraction index distribution:

$$n(\rho) = \frac{2n_0}{1+(\rho/a)^2} \tag{1}$$

where $\rho$ is the distance to the origin, which in the 2D case is $\rho^2 = x^2+y^2$. Within the Geometrical Optics framework, the rays emitted from an arbitrary object point $(x_0, 0)$ are focused onto its image point $(-a^2/x_0, 0)$. Leonhardt found that this perfect focusing property also holds in z-polarised waves [6]. The wave propagating from object point $(x_0, 0)$ into image point $(-a^2/x_0, 0)$, which has been called Leonhard's forward wave [18], is given by:

$$\mathbf{E}(x,y,z) = E_z(x,y)\mathbf{z} = A\left(P_\nu(\zeta) + i\frac{2}{\pi}Q_\nu(\zeta)\right)\mathbf{z}$$

$$\left(\frac{a}{n_o}k_0\right)^2 = \nu(\nu+1) \qquad k_0 = 2\pi f\sqrt{\mu_0\varepsilon_0} \tag{2}$$

$$\zeta = \frac{|z'|^2 - a^2}{|z'|^2 + a^2} \qquad z' = \frac{z-x_0}{z\frac{x_0}{a^2}+1} \qquad z = x+iy$$

Here $A$ is a complex constant, and $P_\nu$, $Q_\nu$ are the Legendre functions [23]. The field $E_z$ diverges at the points $(x_0, 0)$ and $(-a^2/x_0, 0)$, i.e., at the source and at the drain. This solution requires a perfect point drain which is a theoretical concept that can be modeled as an infinitely small region centered around the drain point and with a particular complex permittivity distribution [18].

Using Transformation Optics it was proven [24] that the fields given by Eq. (2) in a 2D MFE are transformed into radial fields in the SGW filled with a refractive index medium with law $n(r) = an_0/r$ (where $r^2 = x^2+y^2+z^2$), see Fig. 1. The radial field $\mathbf{E}(r,\theta,\phi)=E_r(r,\theta,\phi)\mathbf{r}$ in the SGW is related to the MFE field (Eq. (2)) of the corresponding point by:

$$E_r(r,\theta,\phi) = E_z(\zeta) \qquad \zeta = \cos\theta \tag{3}$$

Corresponding points in the MFE and SGW are related by a stereographic projection. Source and drain points $(x_0,0)$ and $(-a^2/x_0,0)$ are transformed into opposite poles of the SGW. Leonhardt's forward wave is transformed into a wave with rotational symmetry with respect to the line passing through object and image points, as shown in Fig. 1. The perfect point drain complex permittivity distribution is transformed into another one.

When the drain of the SGW is not perfect but still rotationally symmetric, the field can be expressed as follows [18]:

$$E_r(r,\theta,\phi) = E_r(\zeta) = AF_\nu(\zeta) + BR_\nu(\zeta) \tag{4}$$

where

$$F_\nu(\zeta) = P_\nu(\zeta) + i\frac{2}{\pi}Q_\nu(\zeta)$$
$$R_\nu(\zeta) = P_\nu(\zeta) - i\frac{2}{\pi}Q_\nu(\zeta) \tag{5}$$

$F_\nu$ is Leonhardt's forward wave and $R_\nu$ can be interpreted as the reverse wave [18]. $\zeta$ is given in Eq. (3).

## 3. Microwave circuit and parameters of the simulation.

The SGW is bounded by two spherical shells made of conductors. The media between shells has the refractive index distribution proportional to $1/r$, as said before. Two coaxial probes loaded with their characteristic impedance have been used to simulate the source and drain in the SGW. The microwave prototype described in [19] and [20] was done in the same way. We shall call them source port and drain port respectively. This drain, when located at the image point does not perform as perfect, that is, it causes a reversed wave ($B \neq 0$ in Eq.(4)). Whilst this implies that there is no perfect focusing as defined before (i.e. full absorption of the forward wave). It does not imply however that perfect imaging cannot occur. This has practical interest since allows the implementation of perfect drains to be avoided.

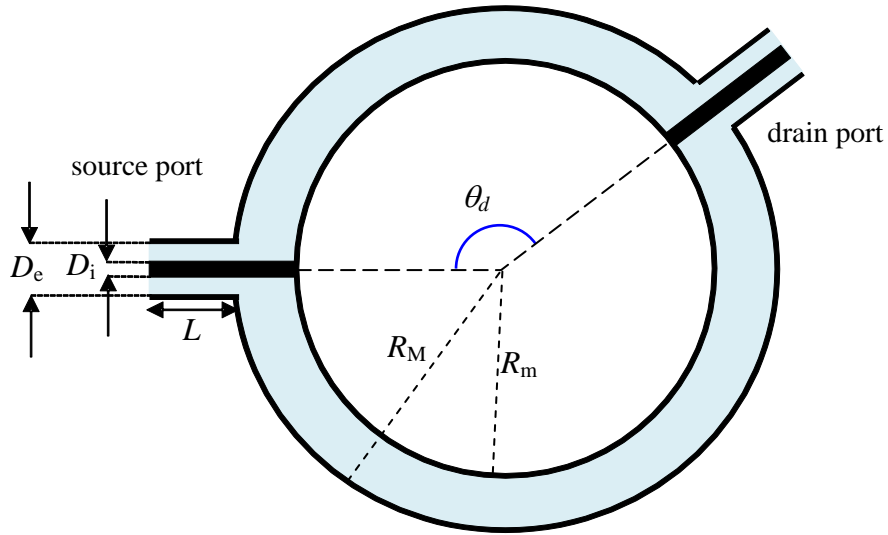

Fig. 2. Cross section of the two coaxial lines and the spherical waveguide (SGW). The power is injected through the source port, The radiation is guided between spheres and may be extracted at the drain port. Both ports are geometrically identical. The drain port is loaded with its characteristic impedance $Z_o$.

Fig. 2 shows the cross section of the SGW with the two coaxial probes simulating the point source and the point drain. The radiation is injected through the source port, guided between spheres, and may (or not) be extracted from the sphere through the drain port. When the angle $\theta_d = \pi$, the drain port is located at the image point and the fields will have rotational symmetry. For $\theta_d \neq \pi$, the rotational symmetry is broken.

The frequencies used in the analysis are low enough so only TEM modes propagate in the coaxial cables. Therefore, the complete system can be analyzed as a microwave circuit using the classical scattering matrix **S** [26]. Fig. 3 shows the equivalent circuit. The definition of the different electrical variables is as follows:

- $V_s^+$, $I_s^+$: voltage and current waves in the source port propagating towards the SGW.
- $V_s^-$, $I_s^-$: voltage and current waves in the source port propagating from the SGW.
- $V_d^+$, $I_d^+$: voltage and current waves in the drain port propagating towards the SGW.
- $V_d^-$, $I_d^-$: voltage and current waves in the drain port propagating from the SGW.
- $Z_0$: characteristic impedance of the coaxial lines.

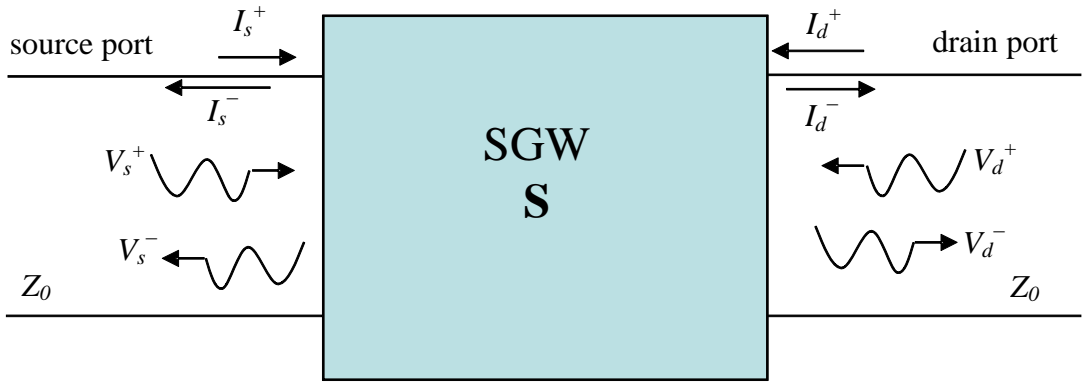

Fig. 3. Microwave circuit made up of the two ports and the spherical waveguide. The sphere S completely characterized with the **S** matrix.

The matrix **S** of the sphere is defined as follows:

$$\begin{bmatrix} V_s^- \\ V_d^- \end{bmatrix} = \begin{bmatrix} S_{11} & S_{12} \\ S_{21} & S_{22} \end{bmatrix} \begin{bmatrix} V_s^+ \\ V_d^+ \end{bmatrix} \qquad (6)$$

When the drain port is matched with its characteristic impedance $Z_o$, there is no reflected wave in this coaxial line and thus the voltage and current waves $V_d^+$ and $I_d^+$ are null:

$$V_d^- = S_{21} V_s^+ \qquad V_s^- = S_{11} V_s^+ \qquad (7)$$

The power injected through the source port $P_I$, transmitted to the drain port $P_T$ and reflected by it $P_R$ are:

$$P_I = \frac{1}{2} \frac{|V_s^+|^2}{Z_o} \qquad P_T = P_I |S_{21}|^2 \qquad P_R = P_I (1 - |S_{21}|^2) \qquad (8)$$

When the drain port is located at the source's image point, and it is loaded with its characteristic impedance, it does not behaves as the perfect drain designed in [18], because with this load the electric field $E_r(\theta)$ is given by Eq. (4) with $A \neq 0$ and $B \neq 0$, i.e. both forward and reverse waves exist.

The simulation has been made using COMSOL and CST Microwave Studio with the following geometrical parameters (see Fig. 2):

$$D_e = 10 \text{ mm} \quad D_i = 5 \text{ mm} \quad L = 20 \text{ mm} \quad R_M = 1005 \text{ mm} \quad R_m = 1000 \text{ mm} \quad (9)$$

The frequency range being analyzed will be from 0.2 GHz to 0.4 GHz ($\lambda$ between 0.75 m and 1.5 m), well below the cut-off frequency of next higher order mode in the coaxial cables (which $\sim(2c/\pi)/(D_e+D_i)$ = 112.7 GHz [26]).

Both software packages lead to similar results. The refractive index between the spherical shells $n(r) = an_0/r$. By selecting $a=R_M$, and $n_0=1$, we obtain $n(r) = R_M/r$ and from the second row in Eq. (2):

$$(R_M k_0)^2 = \nu(\nu+1) \quad (10)$$

Since $R_M/R_m \approx 1$, we have approximated $n=1$ inside the waveguide for the CST Microwave Studio model. The COMSOL model uses the gradient index material. As said before, there is no significant difference in the results. The port radius is less than $\lambda/100$ for the analyzed frequencies and the ratio $D_e/R_m<<1$ in the aim of properly modeling the point nature of source and drain. The coaxial lines are 20 mm long, which is enough to guarantee the evanescent modes in the coaxial lines are negligible at their ends. Two conductor types have been considered: a perfect conductor and copper. Both lead to nearly identical results.

Special care has been taken to define the mesh of the system. In the CST Microwave Studio we used the auto-scaling option, while in COMSOL we selected user-defined grids. Fig. 4 shows an example of mesh used in COMSOL.

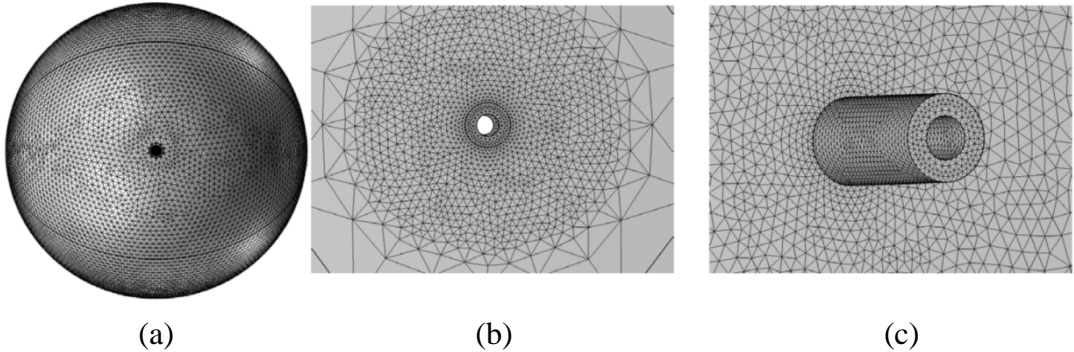

(a)          (b)          (c)

Fig. 4. Mesh structure. (a) SGW, (b) close up of one probe from the inner side of the sphere, (c) close up of one probe from outside the sphere.

## 4. Power transmitted for different drain port positions and different frequencies.

Several simulations have been made to analyze the imaging properties of the system. We have used $|S_{21}|^2$ to determine the sensitivity of the transmitted power $P_T$ (which is proportional to $|S_{21}|^2$, see Eq. (8)) to the drain port position.

### 4.1. $|S_{21}|^2$ as function of frequencies for different drain port positions.

Fig. 5 shows $|S_{21}|^2$ for a frequency range between 0.2 GHz and 0.4 GHz when the drain port is placed at the source's image point, that is, $\theta = 0$ for the source port and $\theta = \pi$ for the drain port. There are peaks of $|S_{21}|^2$ indicating total transmission from the source port towards the drain port, resembling the transmission diagram of a Fabry-Pérot resonator (see for instance [27]). These peaks occur at the well-known Schumann resonance frequencies of the spherical systems [28], which correspond to integer values of $\nu$ in Eq. (10).

Fig. 6 shows $|S_{21}|^2$ when the drain port is shifted $\lambda/30$ (for $\lambda=1$m corresponding to 0.3 GHz)

away of source port antipode. Although the results are extremely similar, narrow notches in the transmission very close to the Schumann frequencies occur. These notches widen when the drain port is shifted further from the source's image point, as can be seen better in the close in of Fig. 7.

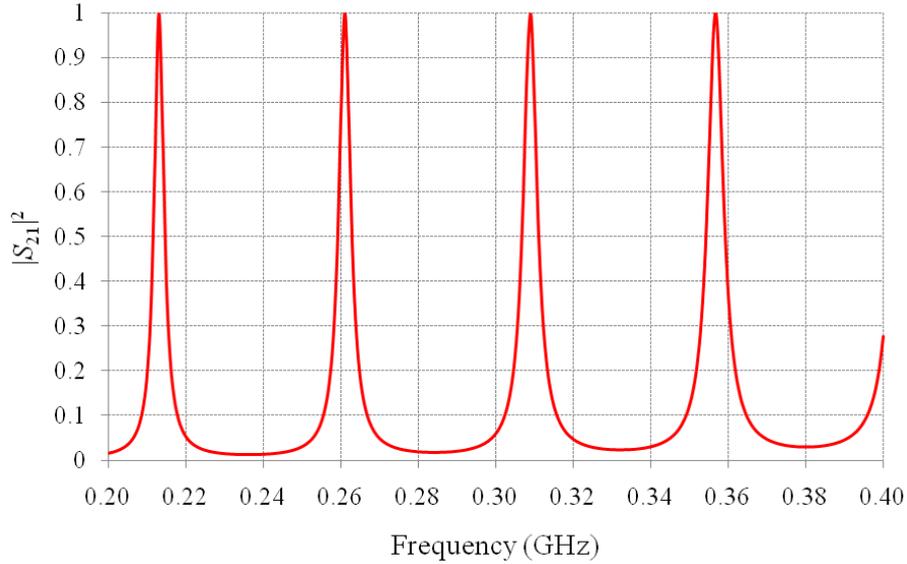

Fig. 5. $|S_{21}|^2$ as function of frequency when the drain and source ports are at opposite poles. The peaks occur at the Schumann resonance frequencies.

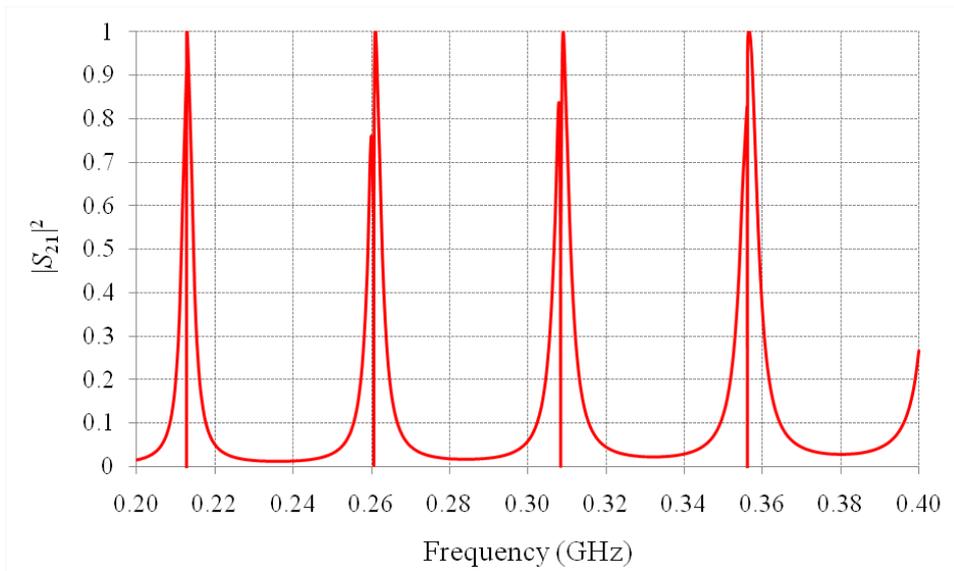

Fig. 6. $|S_{21}|^2$ vs frequency when the drain port is shifted $\lambda/30$ ($\lambda=1$m) from the source port antipode. The results are similar to those presented in Fig. 5 in accordance with the classical prediction, except for the very narrow notches near Schumann frequencies.

Fig. 7 shows $|S_{21}|^2$ for different drain port positions in a very narrow band in the neighborhood of the Schumann frequency corresponding to the second peak in Fig. 6 (for which $\nu=5$). The label of each curve indicates the distance between the drain port center and the source port antipode. The black curve corresponds to the drain port placed in the source port antipode (it looks flat because of the high zoom in the frequency axis). The other curves correspond to different shifts of the drain port. The shifts are in all cases much smaller than wavelength (from $\lambda/33$ to $\lambda/500$ with $\lambda=1.15084047$ m that correspond to $f=0.2606873$ GHz, see Fig. 7). These results are quite surprising, since close to a specific frequency the power transmitted to the drain port suddenly reduces to a value near zero, indicating super-resolution.

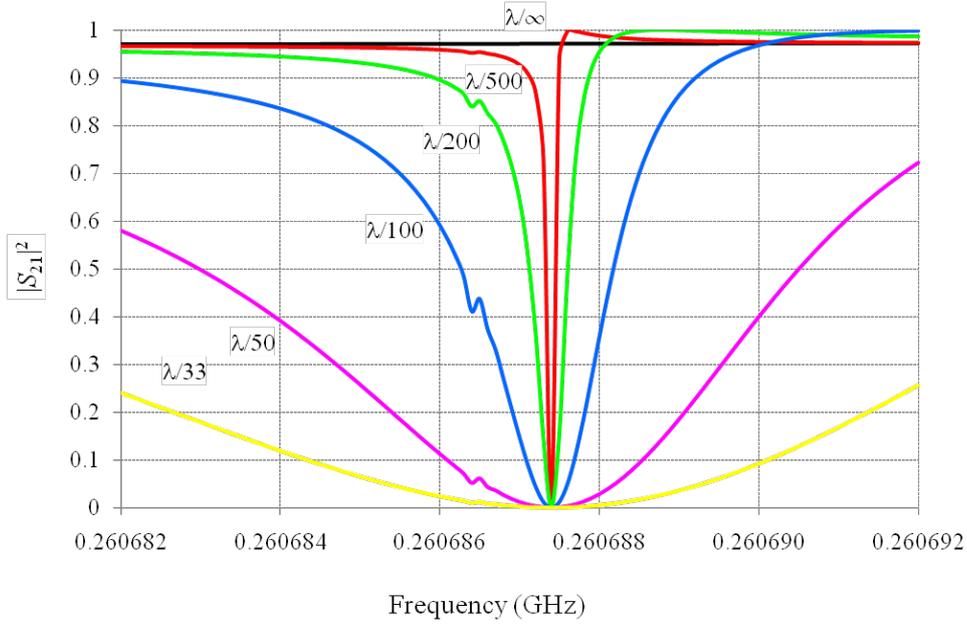

Fig. 7. Detailed picture of $|S_{21}|^2$ as function of the frequency in a narrow band around a Schumann frequency for different drain port positions (the corresponding shift on the inner sphere of the SGW between the drain port centre and the source port antipode has been used for labelling). No shift is labelled as $\lambda/\infty$ and it is the only curve that does not show a notch in the transmission. The notches have a fixed null about f=0.2606873 GHz. The nearest Schumann frecuency is f=0.26086609 GHz which is out the range of this Figure

### 4.2. $|S_{21}|^2$ as function of drain port shift for different frequencies.

Fig. 8 shows the same information as Fig. 7 but plotting $|S_{21}|^2$ vs the drain port shift (expressed in units of $\lambda$) and using the frequency as a parameter. All the frequencies chosen are slightly above the Schumann frequency (graphs for frequencies slightly below are quite similar). Since $|S_{21}|^2$ is proportional to the transmitted power, this Figure is somehow equivalent to the Point-Spread-Function commonly used in Optics, and it is therefore adequate to show the super-resolution properties of the SGW for different frequencies.

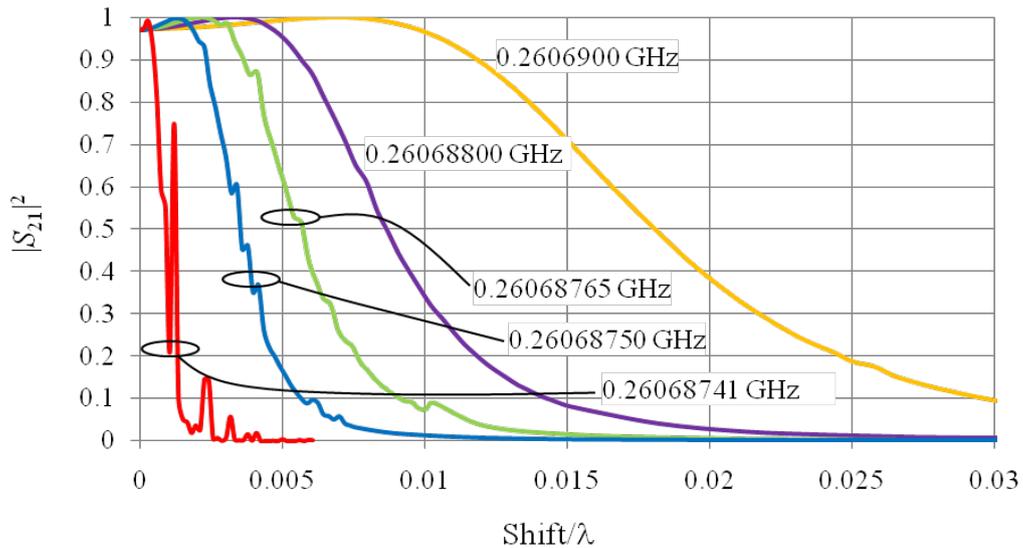

Fig. 8. $|S_{21}|^2$ as function of the drain port shift for different frequencies that present super-resolution of between $\lambda/30$ and $\lambda/500$.

Let us define "resolution" as the arc length that drain port needs to be shifted so $|S_{21}|^2$ drops to

10% (not far from Rayleigh criteria in Optics, which refers to the first null). From the orange to the red curves, increasing super-resolutions are achieved: 0.03 $\lambda$ (that is, $\lambda/33$) for the orange to $\lambda/500$ for the red. The latter, whose frequency $f=0.26068741$ GHz corresponds to $\nu=4.99636$) is the highest resolution that we have obtained. Computations for frequencies closer $\nu=5$ show essentially null $|S_{21}|^2$ values for shifts $> \lambda/500$ (as in the red line in the picture), but also instabilities in the $|S_{21}|^2$ values for shifts below $\lambda/500$.

It seems that Leonhardt's assertion of infinite resolution (i.e., perfect imaging) occurs for the discrete Schumann frequencies in the SGW, although the aforementioned instabilities have prevented us from proving it.

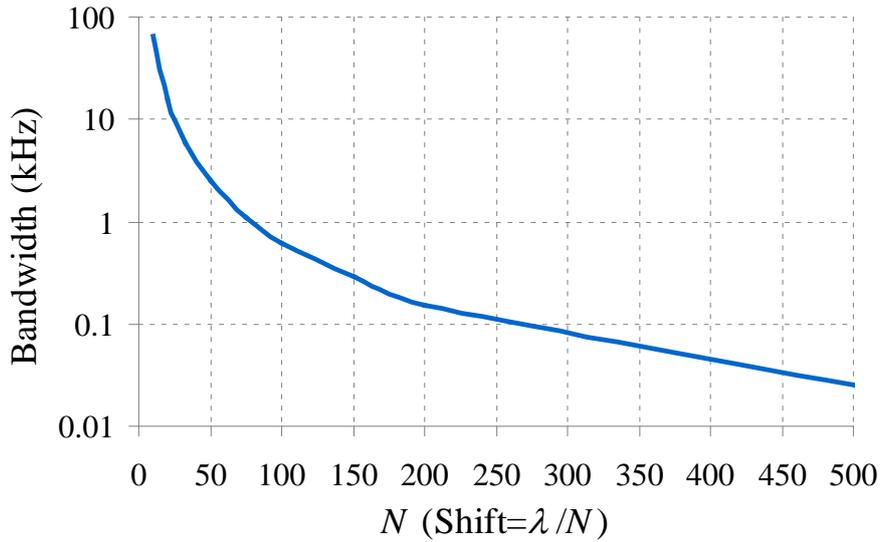

Fig. 9. Bandwidth as a function of the resolution. The ordinate axis shows $N$, meaning that the resolution is better than $\lambda/N$.

The $\lambda/500$ resolution is achieved inside a narrow band (width $\approx 20$ Hz). If larger bandwidths are needed, lower resolutions (but still sub-wavelength) are achieved. Fig. 9 shows the bandwidth vs. the resolution measured with $N$, meaning that the resolution is better that $\lambda/N$. The bandwith has been calculated as $f_{max} - f_{min}$ with $f_{max}$ and $f_{min}$ fulfilling $|S_{21}(f_{max})|^2 = |S_{21}(f_{min})|^2 = 0.1$.

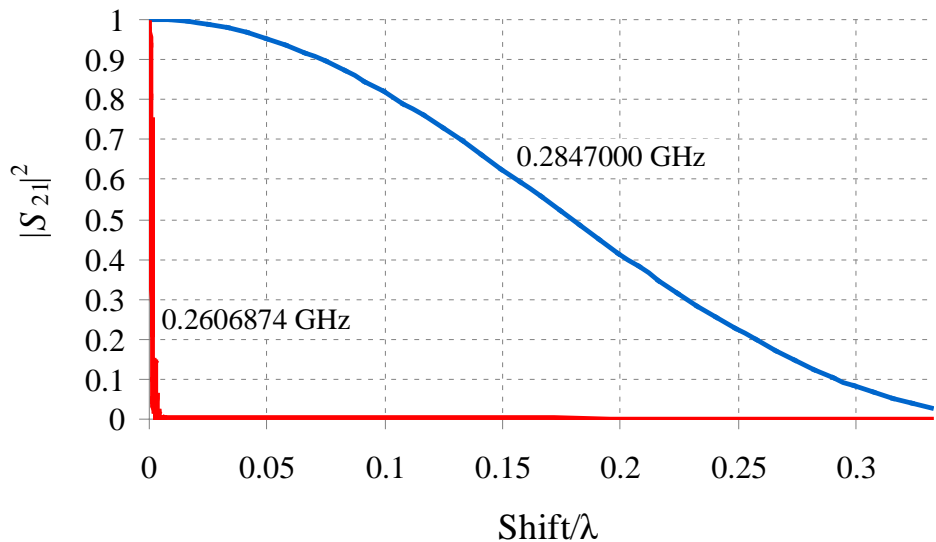

Fig. 10. $|S_{21}|^2$ as function of the drain port shift for a frequency near a Schumann one (red curve) and for a frequency far from a Schumann one (blue curve).

Far from a Schumann frequency, no super-resolution is observed. Fig. 10 shows $|S_{21}|^2$ versus the drain port shift as does Fig. 8 but with a different scale for the abscissa axis, and normalized to value of $|S_{21}|^2$ when the drain port is at source's antipode (see Fig. 5). The red curve, which has λ/500 resolution, corresponds to $f$=0.26068741 GHz ($\nu$ =4.996) is also the red curve in Fig. 8). The blue curve corresponds to a frequency $f$=0.2847 GHz, i.e., far from a Schumann frequency ($\nu$ = 5.5). This blue curve fits with the classical diffraction limit, with a first null at 0.38λ. It is remarkable that this curve is indistinguishable from the energy density distribution in the SGW when there is no drain port (see [9], [19] and Annex in [18]). This energy density distribution is proportional to the function $(P_\nu(\cos\theta))^2$, ($\theta R_{min}$ is equal to the shift). This case corresponds in the MFE to the absence of drain [18], which leads to $A=B$ in Eq. (4).

## 5. Discussion.

The traditional diffraction limit is obtained in the analysis of focal regions in free-space. Although in practical applications a detector is placed in the focal plane, it was implicitly assumed that the limit stands even if the detector is perturbing the field, probably considering that such a perturbation will be small. However, Leonhardt in [6] suggested the possibility of surpassing that limit if the field is strongly perturbed by the detector, when it acts as a perfect drain. Here, computer simulations have shown that super-resolution effectively occurs in the SGW (even when the drain is not perfect, i.e. when there is a wave reflected at the drain), if the frequency is close to the Schumann frequencies, but far from them, even though the drain perturbs the field, the classical resolution limit applies.

The analysis has also shown that the detector perturbs the field in different ways depending on the frequency and the drain port shift. This is shown in the figures of Table 1.

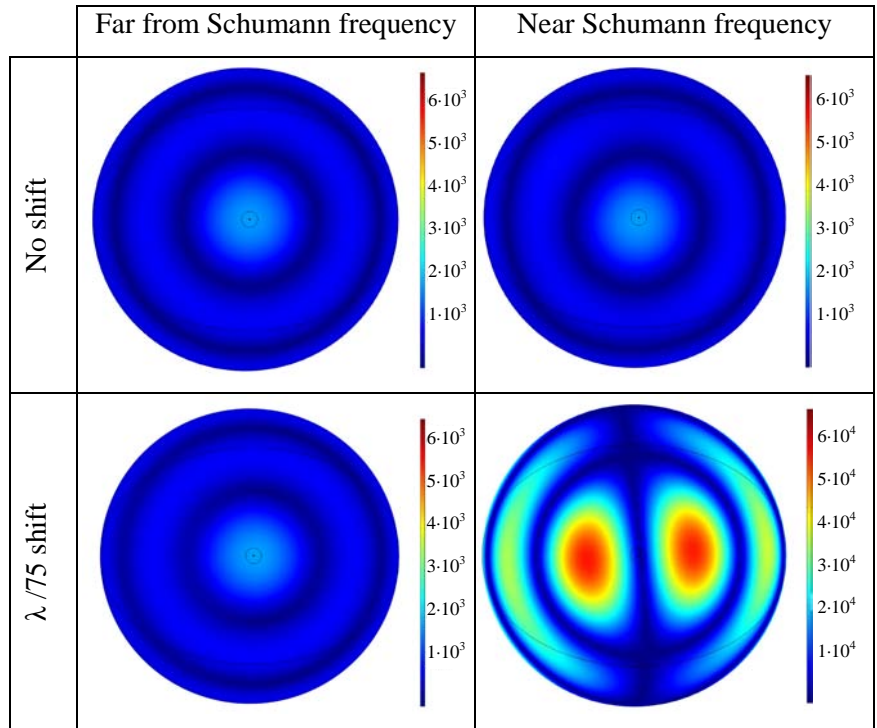

Table 1. Electric field module (in V/m) for 2 shifts of the drain port and 2 frequencies (one far from the Schumann frequency (left) and one near it showing super-resolution. In the pictures the field is expressed in V/m. Incident power of the TEM mode at the source port is 1 W.

Transformation Optics theory showed that the 2D MFE system and the SGW are equivalent for the fields showing super-resolution [24], and so we conclude that super-resolution in the MFE with the drains considered here only occurs near Schumann frequencies. Schumann frequencies are those for which $\nu$ is integer (see Eq. (2)). One way to measure how close a frequency is from a Schumann frequency is just by evaluating the fractional part of $\nu$. This variable takes the value

$v_1 = 9.98$ for the microwave experiment referred to in [19][22], i.e., it is quite close to the Schumann frequency corresponding to $v = 10$; while in the optical experiment referred to in [22] the value is $v_2 = 96.78$ which is not that close. According to our definition of resolution, the microwave experiment would get λ/4 (the reported reference is a resolution of λ/5 with a different definition of resolution). We think that the main reason why only the microwave experiment showed super-resolution is because the frequency chosen is very close to a Schumann frequency.

When a perfect drain is placed at sources's image point in the MFE, perfect focusing occurs for any frequency (not only Schumann frequencies) [6], [18]. However, the impedance that we used as a load, which is the characteristic impedance of the coaxial cable, does not make the drain port a perfect drain. This can be achieved with a specific load impedance that causes the reverse wave inside the SGW (*B*=0 in Eq.(4)) to dissapear. Several questions remain open and are outside the scope of this paper. In particular, what is this impedance necessary to simulate a perfect point drain?; would that impedance provide super-resolution for all frequencies or would the Schumann resonance frequencies again have a special role?; what happens when multiple source and drain ports are used?.

## 6. Conclusions.

Simulations of the spherical waveguide (SGW) showing super-resolution up to $\lambda/500$ at microwave frequencies have been presented. Simulations presented here prove that super-resolution exists in SGW and MFE lens within a narrow band around discrete frequencies as shown in Fig. 6 and Fig. 7. These frequencies are known as Schumann frequencies. Two experimental prototypes have been recently manufactured in order to prove the super-resolution property of MFE. Our results are sound with the two experimental set ups, because the prototype that has shown that super-resolution was tested for a frequency very close to a Schumann one while the other prototype which did not show up any super-resolution response was tested at a non- Schumann frequency.

Although perfect focusing and super-resolution are two strongly linked concepts, they refer to different physical phenomena. Perfect focusing occurs in the MFE and the SGW using perfect drains for an arbitrary frequency ($v$ is an arbitrary positive real number, not necessarily an integer) if they are located at the image point of the source. However, at least with the drains considered here, super-resolution only occurs close to the discrete set of Schumann frequencies ($v$ being an integer). This demonstrates that perfect drains are not necessary to obtain super-resolution.

**Acknowledgments:**

The authors thank the Spanish Ministry MCEI (Consolider program CSD2008-00066, DEFFIO: TEC2008-03773),) for the support given in the preparation of the present work.